%% file: wha.tex
\documentstyle[12pt,epsfig,epsf]{article}
\begin{document}

\newcommand{\nn}{\noindent}
\renewcommand{\thefootnote}{\fnsymbol{footnote}}
\input fey1

\begin{flushright}
IFIC/98/86 \\
FTUV/98/87 \\
FISIST/14-98/CFIF\\
UFR-HEP/98/10 \\
November 1998 \\
\end{flushright}

\vspace*{.4cm}
\begin{center}
{\large\sc {\bf Yukawa coupling corrections to  }}\\
\vspace*{3mm}
{\large\sc {\bf the decay $H^+ \to W^+ A^0$ }} 
\vspace{.4cm}

{{\sc A. Akeroyd}$^{\mbox{a,}}$\footnote{E--mail: akeroyd@flamenco.ific.uv.es},  {\sc A. Arhrib}$^{\mbox{b,c,d,}}$\footnote{E--mail: arhrib@fstt.ac.ma} and 
{\sc E. Naimi}$^{\mbox{c,d}}$}

\vspace{.6cm}
a: Departamento de F\'\i sica Te\'orica, IFIC/CSIC,\\
 Universidad de Valencia, Burjassot 46100,\\
Valencia, Spain

\vspace{.6cm}
b: Departemento de Fisica--CENTRA, Instituto Superior T\'ecnico\\
 Av Rovisco Pais 1, 1096 Lisboa Codex, Portugal

\vspace{.6cm}
c: D\'epartement de Math\'ematiques, Facult\'e des Sciences et Techniques\\
B.P 416, Tanger, Morocco

\vspace{.6cm}

d: UFR--High Energy Physics, Physics Departement, Faculty of Sciences\\
PO Box 1014, Rabat--Morocco

\end{center}

\setcounter{footnote}{4} 
\vspace{1.2cm}

\begin{abstract}
\nn We compute the fermionic radiative contributions to the decay
$H^+ \to W^{+(*)} A^0$ in the framework of models with
two Higgs doublets (2HDM), for
the case of an on--shell and off--shell W. We show that, in the 
majority of the cases, current measurements
of the $\rho$ parameter suggest $M_{H^{\pm}}\ge M_A$
and such decays could invalidate current charged Higgs searches or aid
 detection in the region $M_{H^{\pm}}\approx M_W$. We find that
 the radiative corrections may approach $50\%$ for small values of
 $\tan\beta$. 
\end{abstract}

\newpage

\pagestyle{plain}
\renewcommand{\thefootnote}{\arabic{footnote} }
\setcounter{footnote}{0}

\section*{1.~Introduction}
The search for the Higgs boson ($\phi^0$)\cite{Higgs} of the 
Standard Model (SM)\cite{Wein} 
is one of the major challenges for present and future colliders. In recent
years there has been growing interest in the study of extended Higgs
sectors with more than one Higgs doublet \cite{Gun}. The simplest extension 
is the
Two Higgs Doublet Model (2HDM), and such a structure is required for the
Minimal Supersymmetric Standard Model (MSSM). Models with two (or more)
 Higgs doublets
predict the existence of charged Higgs bosons, and their
discovery would be conclusive evidence of an extended Higgs sector.
In the 2HDM extension of the SM, from the 8 degrees of freedom
initially present in the 2 Higgs doublets, only 5 remain after the electroweak
symmetry breaking and should be manifested as physical particles:
 2 charged Higgs scalars ($H^{\pm}$), 
2 CP--even scalars ($h^0$ and $H^0$) and one CP--odd scalar ($A^0$).
Accurate predictions for the branching ratios (BR) of these particles are
required in order to facilitate the searches and in this paper we consider
the radiative corrections to the decay $H^{\pm}\to A^0W^{(*)}$. In the
non--supersymmetric 2HDM (hereafter to be called simply 2HDM)
 the masses $M_A$ and $M_{H^{\pm}}$ may be taken as free
parameters and so one may consider both the case of an off--shell and 
on--shell $W$. This is in contrast to the MSSM in which
$M_A$ and $M_{H^{\pm}}$ are correlated and the two body decay is never allowed.
We shall show that current measurements of the $\rho$ parameter
strongly suggest $M_{H^{\pm}}\ge M_A$ for $M_{H^{\pm}}\ge 100$ GeV. 

Recently it has been shown that the decay $H^{\pm}\to A^0W^*$ may be
dominant or even close to $100\%$ in the 2HDM (Model~I)
 over a wide range of parameter space relevant at LEP--II \cite{Ake3body}. 
 This 
would affect current charged Higgs searches at LEP--II \cite{Cha}, \cite{OPAL}
 and the Tevatron \cite{Tev} which only assume the decays
$H^{\pm}\to \tau\nu_{\tau}$ and $cs$. We therefore feel it important
to calculate the fermionic radiative corrections 
to this potentially strong tree--level process. An additional
use of the three--body decay would be the possibility of detection 
in the difficult 
$M_{H^{\pm}}\approx M_W$ region, which is considered marginal if $H^{\pm}$
decays conventionally to two fermions. Although a thorough analysis 
is beyond the scope of this paper, the three--body decay would give rise to
high multiplicity signatures of more than 4 jets, with a possibility of
detection above the strong $WW$ background. 
We note that the 2HDM with the
popular Model~II type structure cannot possess a $H^{\pm}$ in the 
discovery range of
LEP--II due to constraints from $b\to s\gamma$ \cite{Giudice} 
(see also ref.\cite{borzu} which derives the 
lower bound $M_{H^{\pm}}\ge 165 $ GeV), while $H^{\pm}$  
in Model~I avoids such constraints and so may be light.
We note
that it is possible to have the Model II type structure and weaken the
above bound on $M_{H^{\pm}}$ in a 2HDM which relaxes natural flavour
conservation (NFC) \cite{Cheung} or a general model with $N(\ge 3$) doublets 
\cite{Gross}. In this paper we are concerned with the 2HDM which imposes
NFC. Limits
on $M_{H^{\pm}}$ from the Tevatron are $\tan\beta$ dependent since one requires
a significant BR$(t\to H^{+}b$) in order to obtain a visible signal. In the 2HDM
with the Model~II type structure this BR can be significant for small ($\le 1$)
or large ($\ge 40$) values of $\tan\beta$. For the Model~I type structure it is only
possible at low $\tan\beta$.

Current mass bounds from LEP--II for the $A^0$ of the MSSM force $M_{H^{\pm}}\ge 
110$ GeV in this model, thus taking $H^{\pm}$ out of the LEP--II discovery range
 \cite{Roy}. In addition, a recent analysis
of the MSSM charged Higgs contributions to $b\to s\gamma$ \cite{Diaz} requires
$M_{H^{\pm}}\ge 110$ GeV, a limit valid in both the MSSM and its simplest
 extension by adding a Higgs singlet superfield (NMSSM). Therefore from the 
 point of view of charged Higgs phenomenology at LEP--II one may consider the 2HDM
(Model~I) but not more popular extended structures.
We will present results for the case of $W$ on--shell and off--shell for
charged Higgs masses of interest at LEP--II and the LHC.
 The paper is organized 
as follows. In Section 2 we introduce our notation and the models in question.
In Section 3 we evaluate the fermionic one loop corrections for the
case of an on--shell and off--shell W, while Section 4 displays the
counterterms. In Section 5 we present our results, and Section 6 contains
our conclusions.

\renewcommand{\theequation}{2.\arabic{equation}}
\setcounter{equation}{0}
\section*{2. Notation, couplings and 
lowest order results}
\subsection*{2.1 Notation and relevant couplings}
In this paper we will use the following notation and conventions.
The momentum of the charged
Higgs boson $H^+$ is denoted by $p_H$ ($p_H$ is incoming), 
${p_W}$ is the momentum of the $W^+$ gauge boson and $p_A$ the 
momentum of the CP-odd $A^0$ ($p_W$ and $p_A$ are outgoing). 
\par
The relevant part of the lagrangian describing the interaction 
of the $W^\pm$ with $H^\pm$ and $A^0$, comes from the covariant
derivative and is given by:
\begin{eqnarray}
{\cal L} = \frac{e}{2 s_W} W_{\mu}^+ 
( H^- \stackrel{\leftrightarrow}{\partial}^{\mu} A^0) + \mbox{h.c.} \label{lag}
\end{eqnarray}
This interaction is model independent (SUSY or non--SUSY) and it depends 
only on standard parameters: electric charge ($e$) and Weinberg angle 
($s_W=\sin\theta_W$).

As we are concerned with the fermionic one loop corrections, we will
give hereafter the relevant couplings.
In the 2HDM there exist four
different ways to couple the Higgs fields to matter
(we assume natural flavour conservation \cite{glashow-weinberg}).
 The two most popular are:
{\sf Model~I:} The quarks and leptons  couple
 only to one of the 2 Higgs doublet exactly as in the minimal
standard model.  {\sf Model~II:} To avoid the problem of Flavor Changing
Neutral Currents (FCNC), one
assumes that one of the 2 Higgs fields 
couples only to down quarks (and charged leptons)
and the other one couples to up quarks (and neutral leptons). 
Model type II is the  pattern found in the MSSM.

In general, the couplings of the charged Higgs boson $H^\pm$,
 Goldstone $G^\pm$, CP--odd $A^0$ and the gauge boson
$W^\pm$  to a pair of fermions are:
\begin{eqnarray}
& & H^+ u\bar d= Y_{ud}^L
\frac{(1-\gamma_5)}{2} + Y_{ud}^R \frac{(1+\gamma_5)}{2}
\qquad , \qquad G^+  u \bar{d} =  G_{ud}^L \frac{(1-\gamma_5)}{2} +
G_{ud}^R \frac{(1+\gamma_5)}{2} \nonumber \\ 
& & A^0 u\bar{u}=  Y_{uu} \gamma_5 \qquad , \qquad 
A^0 d\bar{d}=  Y_{dd}  \gamma_5 \qquad , \qquad
W_{\mu}^+ u\bar{d} = -i\frac{g V_{ud}}{\sqrt{2}}
\gamma_{\mu}\frac{(1-\gamma_5)}{2}
\end{eqnarray}
Where:
\begin{eqnarray}
& & Y_{ud}^L=\frac{g V_{ud} m_u}{ \sqrt{2} M_W 
\tan\beta} \qquad
, \qquad 
 Y_{ud}^R= -\frac{g V_{ud} m_d }{ \sqrt{2} M_W \tan\beta}  \quad 
\mbox{ Model I} \nonumber \\             
&  & Y_{ud}^L=\frac{g V_{ud} m_u}{ \sqrt{2} M_W \tan\beta} \qquad , 
\qquad
Y_{ud}^R=  \frac{g V_{ud} m_d \tan\beta}{ \sqrt{2} M_W} \quad
\ \ \mbox{ Model II} \quad \nonumber\\
& & Y_{uu}= -\frac{gm_u}{2 M_W\tan\beta}\ \ \  \quad , \quad \ \ \
Y_{dd}= \frac{gm_d}{2  M_W \tan\beta} \ \ \ \ \ \quad \mbox{ Model I}  
\nonumber \\
& & Y_{uu}= -\frac{gm_u}{2M_W\tan\beta} \quad \ \ \  , \ \ \  \quad
Y_{dd}= -\frac{m_d g\tan\beta }{2  M_W}\quad \ \ \ \mbox{ Model II}  
\nonumber \\   & & G_{ud}^L =\frac{gm_uV_{ud}}{\sqrt{2}  M_W}
 \qquad\ \ \ \ \ \ , \ \qquad
 G_{ud}^R =-\frac{gm_d V_{ud}}{\sqrt{2} M_W }
\end{eqnarray}
$V_{ud}$ is the Kobayashi--Maskawa matrix element.
It is worth noting that Models I and II are not very different for the 
top--bottom loop
corrections at low $\tan\beta$ because the term $m_t/\tan\beta$ will dominate 
and it is common to both types. 

\subsection*{2.2 Lowest order results}
The lowest--order Feynman diagram for the two body decay 
$H^+\to A^0 W^+$ and
for the three body decay $H^+\to A^0 W^* \to A^0 ff'$ 
are depicted in the following figure:\
\begin{center}
\begin{picture}(100,100)(0,0)
\put(-80,50) {\unbs{$W^+$}}
\put(-80,50) {\unfsp{$A^0$}}
\multiput(-129,50)(8,0){7}{\line(1,0){6}}
\put(-125,54){\mbox{$H^+$}}
\put(-76,50){\circle*{5}}
\put(120,50) {\unbs{$W^+$}}
\put(120,50) {\unfsp{$A^0$}}
\multiput(69,50)(8,0){7}{\line(1,0){6}}
\put(65,54){\mbox{$H^+$}}
\put(124,50){\circle*{5}}
\put(171,6){\line(1,1){24}}
\put(171,5){\line(1,-1){24}}
\put(172,6){\circle*{5}}
\put(195,17){\mbox{$f$}}
\put(195,-12){\mbox{$f'$}}
\put(181.95,17){\vector(-1,-1){1}}
\put(183,-7){\vector(1,-1){1}}
\put(-82,-12){\mbox{Fig1.a}}
\put(118,-12){\mbox{Fig1.b}}
\end{picture}
\end{center}
\vspace{.7cm}
{\centerline{\bf Figure. 1}}
\vspace{.6cm}
In the Born approximation, the decay amplitude of the charged Higgs into 
\underline{on--shell} CP--odd Higgs
boson $A^0$ and the gauge boson $W^+$ (Fig.1.a) can be written as:
\begin{eqnarray}
& & {\cal M}^0 (H^+ \to W^+ A^0 ) = 
 \epsilon^*_{\mu} \Gamma_0^{\mu} \label{amp0}
\qquad \mbox{where} \qquad
  \Gamma_0^{\mu} = i\frac{e}{2 s_W} (p_H + p_A)_{\mu}
\end{eqnarray}
Here $\epsilon$ is the W polarization vector. We then have the following decay width:
\begin{eqnarray}
& & \Gamma_{on}^0 =\frac{\alpha}{16 s_W^2 M_W^2 M_{H^{\pm}}^3 } 
\lambda^{\frac{3}{2}}(M_{H^{\pm}}^2, M_{A}^2,M_W^2 ) \label{width0}
\end{eqnarray}
where $\lambda(x,y,z)=x^2+y^2+z^2 -2(xy+ xz+ yz)$ 
is the familiar two--body phase space function.
Note that in the MSSM the two--body decay of the charged Higgs boson into $W^+
A^0$ is kinematically not allowed.

Below threshold, and taking into account that the virtual $W^*$ decays into 
a pair of fermions $ff'$ ($f\neq t$) (Fig.1.b) which we will take to be 
massless, the Dalitz plot density
for this three--body decay $H^+ \to A^0 W^{+*} \to A^0 ff' $
is given by \cite{dkz}: 
\begin{eqnarray}
\frac{d\Gamma^0_{off}}{dx_1 dx_2} & = & 
9 \frac{\alpha^2}{32\pi s_W^4 }M_{H^{\pm}} 
\frac{ [ (1-x_1)(1-x_2) - \kappa_A ]  }{( [1 -x_1 -x_2  - 
\kappa_A  +\kappa_W]^2 + \kappa_W \gamma_W ) }\nonumber 
\end{eqnarray}
where 
$$ \kappa_{A,W}=\frac{M_{A,W}^2}{M_{H^{\pm}}^2} \qquad , \qquad 
\gamma_{W}=\frac{\Gamma_W^2}{M_{H^{\pm}}^2},$$
$\Gamma_W$ is the total width of the W gauge boson and
$x_{i}=2 E_i/M_{H^{\pm}} $ are the scaled energies of the 
massless fermions in the final state. We note that in the non--SUSY 2HDM
null--searches at LEP in the $e^+e^-\to h^0A^0, h^0Z$ channels
eliminate regions in the $M_A,M_h$ plane \cite{OPAL}, \cite{hAsearch}.
The excluded region does not have a simple shape, and there are
still areas which allow $M_A+M_h\le 90$ GeV. Thus
$M_A$ may be taken as light as $10$ GeV. This is in contrast to the MSSM
in which one can derive individual lower limits on the masses,
of $M_h\ge 70.7$ GeV and $M_A\ge 71.0$ GeV
\cite{hAsearch}. Therefore the 
off--shell
decay in the 2HDM can be relevant even for a small $M_{H^{\pm}}$ ($\le 80$
 GeV) in range at LEP--II.

\renewcommand{\theequation}{3.\arabic{equation}}
\setcounter{equation}{0}

\section*{3. Fermionic radiative corrections.}
We have evaluated the fermionic radiative corrections 
to $H^+ \to W^+ A^0$ (for both the on--shell and off--shell $W$)
at the one loop level. This set of  corrections is Ultra--Violet (UV)  divergent. 
The UV singularities are treated by  dimensional
regularization \cite{thooft} in the on--mass--shell renormalization scheme.\\
The typical Feynman diagrams for the virtual corrections of order $\alpha$
are drawn in figure 2. These comprise the vertex correction 
(Fig.2.$a_1$, Fig.2.$a_2$ ), 
$W^+$--$W^-$ 
self--energy (Fig.2.$a_3$) and the mixed $W^+$--$G^-$ 
self--energy (Fig.2.$a_4$).
Note that diagrams 2.$a_3$ and 2.$a_4$ are not to be considered if 
the gauge boson $W$ is on-shell. These contributions 
have to be supplemented by 
the counterterm renormalizing the vertex $H^+A^0W^- $ (Fig.2.$c_1$),
the counterterm for the off--shell $W$ gauge boson 
self--energy (Fig.2.$c_2$) and 
by the counterterm for the mixing W--G (Fig 2.$c_3$).
These Feynman diagrams are generated and computed 
using FeynArts and FeynCalc \cite{seep,rolf} packages. We also use  the fortran FF--package \cite{ff} in the numerical analysis.
Note that in the general 2HDM, the vertices $W^+A^0 G^-$, 
$W^+G^0 H^-$  and $A^0 H^+ H^-$ are not present,
and so the mixing $G^+$--$H^-$, $G^0$--$A^0$ and $W^+$--$H^-$ does
 not give any  contribution to our process.

The one loop  amplitude  ${\cal M}^1$  can be written as:
\begin{eqnarray}
& & {\cal M}^1 (H^+ \to W^+ A^0 ) =  \epsilon^*_{\mu} \Gamma^{\mu} 
\label{amp1}
\end{eqnarray}
Using Lorentz invariance, $\Gamma^{\mu}$ can be projected as:
\begin{equation}
\Gamma^{\mu}  =\frac{e}{2 s_W}(\Gamma_{H} p_H^{\mu} +\Gamma_{W} p_W^{\mu}) 
\end{equation}
$\Gamma_{H}$ and $\Gamma_{W}$ can be cast as follow:
\begin{eqnarray}
 & & \Gamma_{W}= \Gamma_{W}^{vertex} +\Gamma_{W}^{W^+W^-}+
\Gamma_{W}^{W^+G^-} +
 \delta\Gamma_{W}^{vertex} 
+  \delta\Gamma_{W}^{W^+W^-} + \delta\Gamma_{W}^{W^+G^-} \\
 & & \Gamma_{H}= \Gamma_{H}^{vertex} +  \Gamma_{H}^{W^+W^-} + 
\delta\Gamma_{H}^{vertex} 
+  \delta\Gamma_{H}^{W^+W^-}\label{forms}
\end{eqnarray}
Where $\Gamma_{W,H}^{vertex}$, $\Gamma_{W,H}^{W^+W^-}$  
and $\Gamma_{W}^{W^+G^-}$
are respectively the contribution of the two vertices, 
 the contribution of the self--energy of the $W$ and
 the contribution of the mixed $W^+G^-$ self--energy; 
$\delta\Gamma_{W,H}^{vertex}$, $\delta\Gamma_{W,H}^{W^+W^-}$  and
$\delta\Gamma_{W}^{W^+G^-}$ are the counterterms needed to remove the
 UV divergences
contained in $\Gamma_{W,H}^{vertex}$, $\Gamma_{W,H}^{W^+W^-}$  and 
$\Gamma_{W}^{W^+G^-}$. In what follows, we write the above 
one loop corrections explicitly. The expressions for the counterterms
can be found in Section 4.

\subsection*{3.1 Vertex with u--u--d exchange: Fig.2.$a_1$}
The  amplitude of the u--u--d quarks contribution 
to $H^+ A^0 W^+$ vertex  is given by
\begin{eqnarray}
& & \Gamma_{H}^{uud} =N_C\frac{  \alpha}{2 \pi \sqrt{2} M_{H^{\pm}}^2} 
Y_{uu} \Bigm (  
(-m_D^2 -3 M_{H^{\pm}}^2 + m_U^2) Y_{ud}^L B_0(M_{H^{\pm}}^2 ,m_D^2, m_U^2)
+ \nonumber\\ & &
(m_D^2 - m_U^2) Y_{ud}^L  B_0(0, m_D^2, m_U^2)  - 
       2 M_{H^{\pm}}^2 \{ (m_U^2 Y_{ud}^L + m_D m_U Y_{ud}^R) 
C_0 + \label{gamma1} \nonumber\\ & &
          Y_{ud}^L [ p_W^2 C_1 -  2 C_{00} + 
(- M_A^2 + M_{H^{\pm}}^2 -p_W^2) C_{12} - 
          2 M_A^2 C_{2 2} ] \} \Bigm )  \\ \nonumber \\
& & \Gamma_{W}^{uud} = N_C\frac{ \alpha}{\pi \sqrt{2} } 
 Y_{uu} \Bigm ( Y_{ud}^L B_0 (M_{H^{\pm}}^2, m_D^2, m_U^2) + 
       m_U( m_U Y_{ud}^L + m_D Y_{ud}^R)
C_0 - \nonumber\\ & &
       Y_{ud}^L \{ (M_A^2 - M_{H^{\pm}}^2) C_{1} +  M_A^2 C_{2} +  2 C_{00} + 
 (M_A^2 - M_{H^{\pm}}^2 + p_W^2) C_{11} + \nonumber\\ & &
 (3 M_A^2 - M_{H^{\pm}}^2 + p_W^2) C_{12} + 
          2 M_A^2 C_{22} \} \Bigm )
\end{eqnarray}
with $A_0$, $B_0$, $C_i$ and $C_{ij}$ the Passarino-Veltman functions
\cite{passarino} which we define in Appendix A. $N_C=3$ for quarks and 1 for leptons. 
All the $C_i$ and $C_{ij}$ have the same arguments: 
$(p_W^2, M_{H^{\pm}}^2, M_A^2, m_U^2, m_D^2, m_U^2)$

\subsection*{3.2 Vertex with d-d-u exchange: Fig.2.$a_2$}
 The amplitude of this diagram 
can be obtained from the 
above one just by making the following replacement:
\begin{eqnarray}
\Gamma_{H,W}^{ddu}=\Gamma_{H,W}^{uud}
[ m_U \longleftrightarrow m_D \quad , \quad Y_{ud}^R \longleftrightarrow  
Y_{ud}^L\quad ,  \quad Y_{uu}\rightarrow Y_{dd} ] \nonumber
\end{eqnarray}
\noindent
The total contribution of vertex is:
$$\Gamma_{W,H}^{vertex} = \Gamma_{W,H}^{uud} + \Gamma_{W,H}^{ddu} $$

\subsection*{3.3 $W^+$--$W^-$ self--energy: Fig.2.$a_3$}
The contribution of $W$ self--energy Fig.2.$a_3$ evaluates to
\begin{eqnarray}
& & \Gamma_{H}^{WW} = \frac{N_C \alpha }{2 \pi  s_W^2 (p_W^2-M_W^2) }
\{ A_0(m_U^2) + m_D^2 B_0(p_W^2, m_D^2, m_U^2) - 
 2 B_{22}(p_W^2, m_D^2, m_U^2)\nonumber \\ [0.2cm] & &
 + p_W^2 B_1(p_W^2, m_D^2, m_U^2) \} \nonumber \\ 
&  & \Gamma_{W}^{WW} =
- \frac{N_C \alpha }{4 \pi  s_W^2 (p_W^2-M_W^2) }
\{ A_0[m_U^2] + m_D^2 B_0(p_W^2, m_D^2, m_U^2) - 
2 B_{22}(p_W^2, m_D^2, m_U^2) \nonumber \\  [0.2cm] & &
 +  p_W^2 B_1(p_W^2, m_D^2, m_U^2) +
        2 (M_{H^{\pm}}^2 - M_A^2) ( B_1(p_W^2, m_D^2, m_U^2) + 
B_{21} (p_W^2, m_D^2, m_U^2) \}
\end{eqnarray}
\subsection*{3.4 W--G mixing: Fig.2.$a_4$ }
In accordance with Lorentz invariance, the mixing self--energy W--G is 
proportional to $p_W^\mu$ and evaluates to
\begin{eqnarray}
& & \Gamma_{W}^{WG} =
\frac{N_C \alpha (M_{H^{\pm}}^2 - M_A^2) }{4 \pi M_W^2 s_W^2 (p_W^2-M_W^2) }
\{ m_D^2 B_0(p_W^2, m_D^2, m_U^2) + 
       [ m_D^2  - m_U^2] B_1(p_W^2, m_D^2, m_U^2)\}\nonumber \\
& & \Gamma_{H}^{WG}=0
\end{eqnarray}

\renewcommand{\theequation}{4.\arabic{equation}}
\setcounter{equation}{0}

\section*{4. On--mass--shell Renormalization.}
The parameters entering the tree--level amplitude in eq.(\ref{amp0}) are all
standard model parameters ($e$ and $s_W$).
This fact will render the one loop renormalization rather simple,
in the sense that all non--standard parameters appearing first at the
one loop level (like $\tan\beta$), will not get renormalized. This
is in contrast to the calculation in \cite{santos} for the process
$H^+\to hW^+$ which explicitly contains the factor $\cos^2(\beta-\alpha)$
at tree--level. Therefore renormalization conditions related to the
 definition of $\tan \beta$ are not explicitly 
needed here. We will need, however, to renormalize the electric charge,
the Weinberg angle, charged Higgs
wave--function, CP--odd Higgs wave function 
and $W$ gauge boson wave function. In our case the $W^\pm$ gauge boson
 mixes with
the Goldstone boson $G^\pm$, by virtue of the Laurentz invariance
 of the self--energy; therefore
$\Sigma_{\mu}^{G^+ W^-}$ is proportional to $p^W_{\mu}$ and so if the $W$ is
 on-shell
the mixing would have a vanishing contribution but in the off--shell case we 
have to take this mixing into account.

In what follows we will follow to an extent the on--shell--renormalization 
developed by R. Santos et al \cite{santos} which is 
the generalization to the 2HDM of the I. Aoki et al on--shell renormalization 
scheme \cite{aoki, denner}.
The crucial point in this scheme is that all fields and masses
 are renormalized after the
diagonalization of the bare mass matrices.  Another important point
 in this scheme
is that the gauge fixing is  written
in terms of the renormalized parameters and fields and as a consequence
 it does not contain any counterterm.

\subsection*{4.1 Vertex $H^+A^0W^+$ counterterm}
To obtain the renormalized vertex  $W^-A^0 H^+$ vertex we have to make the
following substitutions in eq. (\ref{lag}):
\begin{eqnarray}
& & W_{\mu}  \to Z_W^{1/2} W_{\mu} \label{deltaw} \\
& & H^{\pm}  \to Z_{H^+H^+}^{1/2} H^{\pm} \label{deltah} \\
& & A^0  \to Z_A^{1/2} A^0\nonumber \\ 
& & e \to Z_e e = (1+ \delta Z_e)e \nonumber \\ 
& & M_W^2 \to M_W^2 + \delta M_W^2 \label{dmw2} \\ 
& & M_Z^2 \to M_Z^2 + \delta M_Z^2 \nonumber 
\end{eqnarray}
Note that in the on--shell scheme, the Weinberg angle is defined as:
$s_W^2 = 1-\frac{M_W^2}{M_Z^2}$.
Therefore the counterterm of $s_W$ is completely fixed by 
the counterterm of the W and Z
boson masses and is given by:
\begin{eqnarray}
\frac{\delta s_W}{s_W}  =  -\frac{1}{2} \frac{c_W^2}{s_W^2} 
( \frac{\delta M_W^2}{M_W^2} - \frac{\delta M_Z^2}{M_Z^2} )  
\end{eqnarray}
Setting $Z^{1/2}=1+ \frac{1}{2} \delta Z$, one obtains the following
 counterterm:
\begin{eqnarray}
\delta {\cal L} = \frac{e}{2 s_W} W_{\mu}^+ 
(H^- \stackrel{\leftrightarrow}{\partial}^{\mu} A^0 )( \frac{1}{2} \delta Z_W +
\frac{1}{2} \delta Z_{A^0}+
\frac{1}{2} \delta Z_{H^{\pm}H^{\pm}} + \delta Z_e - \frac{\delta s_W}{s_W} )
 \label{dlag}
\end{eqnarray}

In the on--mass--shell scheme the counterterms can be fixed by the 
following renormalization conditions:
\begin{itemize}
\item On--shell condition for the charged Higgs boson $H^\pm$, CP--odd $A^0$
and the $W$ and $Z$ gauge Bosons.
We choose to identify
the physical  mass with the corresponding parameter in the renormalized 
lagrangian, and require the residue of the propagator to have its tree--level 
value, i.e., 
\begin{eqnarray}
\delta M^2 = Re \, {\Sigma } (M^2)\ \
\mbox{and} \ \ 
 \delta Z = -\frac{\partial\Sigma  (k^2)}{\partial k^2} |_{k^2=M^2}\label{dmz}
\end{eqnarray}
where $\Sigma (k^2)$ is the  bare self--energy of the $H^{\pm}$, 
$A^0$ or  $W$.
\item  the electric charge $e$ is defined as in the minimal standard 
model \cite{denner, Hollik}.
\item Tadpoles are renormalized in such a way that the renormalized tadpoles
vanish: $T_{h} +\delta t_h=0$, $T_H +\delta t_H=0$. These conditions 
guarantee that $v_{1,2}$ appearing in the renormalized lagrangian
 are located at the minimum of the one loop potential.
\end{itemize} 
Using these renormalization conditions and as is shown in \cite{denner},
the renormalization constant of the electric charge and 
 counterterm of gauge boson mass are given by:
\begin{eqnarray}
& & \delta Z_e = -\frac{1}{2} \delta Z_{\gamma\gamma} +
\frac{1}{2} \frac{s_W}{c_W} \delta Z_{Z\gamma} = 
\frac{1}{2} \frac{\partial\Sigma_T^{\gamma\gamma}(k^2)}{\partial k^2}
\vert_{k^2=0}
+ \frac{s_W}{c_W}  \frac{\Sigma_T^{\gamma Z}(0)}{M_Z^2}\label{dze}\\
& & \delta M_W^2 =\Sigma_T^{WW}(M_W^2) \qquad \mbox{and }  \qquad
\delta M_Z^2 =\Sigma_T^{ZZ}(M_Z^2)\label{dmv2}
\end{eqnarray}
Where $\Sigma_T^{WW}$, $\Sigma_T^{ZZ}$ $\Sigma_T^{\gamma\gamma}$
are  respectively  the W, Z and  photon self--energies depicted in
Fig.2.$b_{3,4,5}$,  T index  is to denote that we take only the transverse part.
We stress at this stage that the fermionic contribution to the 
 mixing  $\Sigma_T^{\gamma Z}(k^2)$ vanishes at $k^2=0$.

\subsection*{4.2 Counter term for the W self--energy and the mixing W--G}
 One obtains the counterterm for the $W$--$W$ self--energy by substituting
 eqs(\ref{deltaw}, \ref{dmw2}) in the $W$ lagrangian:
\begin{eqnarray}
\delta (W^{\mu} W^{\nu}) = i(g^{\mu\nu} -\frac{p_W^\mu p_W^\nu}{p_W^2} ) 
(\delta M_W^2
+ (M_W^2 - p_W^2) \delta Z_W ) + i\frac{p_W^\mu p_W^\nu}{p_W^2} (\delta M_W^2 +
M_W^2 \delta Z_W )
\end{eqnarray}
All the counterterms appearing in $\delta (W_{\mu} W_{\nu}) $ 
are fixed by the  renormalization conditions fixed 
above eqs. (\ref{dmz}, \ref{dmv2}).

As we have mentioned above, $W^+$ boson and $G^+$ goldstone mix. 
To treat this mixing, R.Santos et al \cite{santos}
have considered the mixing of $G^+$--$H^-$ which they have
renormalized in the following way:
\begin{eqnarray}
& & H^{\pm}  \to Z_{H^+H^+}^{1/2} H^{\pm} + 
Z_{H^+G^+}^{1/2} G^{\pm} \label{hg} \\
& & G^{\pm}  \to Z_{G^+G^+}^{1/2} G^{\pm} + 
Z_{G^+H^+}^{1/2} H^{\pm} \label{gh}
\end{eqnarray}
At the one loop level $Z_{ii}^{1/2}=1+1/2 \delta Z_{ii}$ and 
$Z_{ij}^{1/2}= \delta Z_{ij}$ where $\delta Z_{ij}={\cal O}(\alpha)$. 
These four renormalization constants together with the counterterm mass
of the charged Higgs bosons are fixed by imposing the on--shell condition 
(mass located at
the pole of the propagator and residue equal to one) and the vanishing mixing
both for $\Sigma_{G^+H^+}(k^2)$ self--energy at $k^2=M_{H{\pm}}^2$
and $\Sigma_{H^+G^+}(k^2)$ self--energy at $k^2=0$.
Note that the Goldstone
boson receives its renormalized mass from the gauge fixing lagrangian. Before 
introducing this lagrangian the Goldstone boson is massless, and so the
renormalization 
conditions imposed on the propagator of the Goldstone and its mixing
 with charged Higgs boson will be fixed at $k^2=0$.

At the one loop level the renormalization constants $\delta Z_{H^+H^+}$ and 
$\delta Z_{G^+G^+}$ are given by
\begin{eqnarray}
\delta Z_{H^+H^+} = 
-\frac{\partial {\Sigma}_{H^+H^+} (k^2) }{\partial k^2}|_{k^2=M_{H{\pm}}^2} 
\qquad \mbox{and } \qquad
\delta Z_{G^+G^+} = 
-\frac{\partial{\Sigma}_{G^+G^+} (k^2) }
{\partial k^2} |_{k^2=0}
\end{eqnarray}
Performing the replacement (\ref{deltaw}, \ref{dmw2} and \ref{gh}) in 
the W gauge fixing term $i M_W \partial^{\mu} W^+_{\mu} G^-$, 
generated from the  covariant derivative, one finds the following 
counterterm for the  mixing $W^+$--$G^-$:
\begin{eqnarray}
\delta ( W_{\mu}^+ G^-) = i p_W^{\mu} M_W (\frac{\delta M_W^2}{M_W^2} + 
\frac{1}{2}(\delta Z_W + \delta Z_{G^+ G^+}) )
\end{eqnarray}   
This  completes  the set of counterterms needed for our study. 
The renormalization
constants of the wave function and the mass counterterms are given in
the appendix B.

\subsection*{4.3 Back to counter--terms form factors }  
After the  short discussion in section 4.2 about the on--shell renormalization
we are using, 
we are now able to give the expressions of the counterterms 
$\delta\Gamma_{W,H}^{vertex} ,  \delta\Gamma_{W,H}^{W^+W^-} ,
 \delta\Gamma_{W}^{W^+G^-} $ defined in eq. (\ref{forms})
\begin{eqnarray}
& & \delta\Gamma_{W}^{vertex} = -(\delta Z_e -\frac{\delta s_W}{s_W} +
\frac{1}{2} (\delta Z_{H^+H^+} + \delta Z_{A^0} + \delta Z_{W}) )\nonumber\\
& &   \delta\Gamma_{H}^{vertex} = -2 \delta\Gamma_{W}^{vertex}\nonumber \\
& & \delta\Gamma_{W}^{WW}  = \{ (M_{H^{\pm}}^2 + p_W^2 -m_{A}^2) \delta Z_W
- \delta M_W^2 - M_W^2 \delta Z_W \}/ (p_W^2 - M_W^2)\label{deltas}\\
& & \delta\Gamma_{H}^{WW}  =2 \{ 
 \delta M_W^2 + (M_W^2 -p_W^2) \delta Z_W  \}/ (p_W^2 - M_W^2)\nonumber  \\
& & \delta\Gamma_{W}^{WG}  =\frac{1}{2}(M_{H^{\pm}}^2 -M_A^2) 
\{ \frac{\delta M_W^2}{M_W^2} + \delta Z_W  +
 \delta Z_{G^+G^+}\} /(p_W^2 -M_W^2)\nonumber
\end{eqnarray}

\renewcommand{\theequation}{5.\arabic{equation}}
\setcounter{equation}{0}

\section*{5. Numerical results and discussion}

In the previous section we have summarized the analytical formulae for the 
fermionic ${\cal O} (\alpha)$ radiative correction to the decay 
$H^+ \to W^+ A^0$.
In this section we focus on
the numerical analysis. 
We take the following experimental input for the physical parameters 
\cite{databooklet}:
\begin{itemize}
\item the fine structure constant: $\alpha=\frac{e^2}{4\pi}=1/137.03598$.
\item the gauge boson masses: $M_Z=91.187\ GeV$, $M_W=80.41\ GeV$ and
$\Gamma_W =  2.06\ GeV $
\item the input lepton masses: $
m_e=0.511 \mbox{MeV} \ , \  m_{\mu}=0.1057 \mbox{GeV} \ , \ 
  m_{\tau}=1.784  \mbox{GeV}  $
\item for the light quark masses we use the effective values which are chosen in 
such a way that the experimentally extracted hadronic part of the 
vacuum polarizations is reproduced 
\cite{martinzepenfieldVerzeganssijegerlener}:
\begin{eqnarray}
& &m_d=47\ MeV \ \ \ \ \ \ \ \ \ \ \ \ \ \ \ \ m_u=47 \ MeV 
\ \ \ \ \ \ \ \ \ \ \ \ 
\ \ m_s=150\ MeV \nonumber\\
& &m_c=1.55\ GeV \ \ \ \ \ \ \ \ \ \ \ \ \ \ \ \ m_b=4.5 \ GeV \nonumber
\end{eqnarray} 
\end{itemize}
For the top quark mass we take $m_t=175$ GeV. 
In the on--shell scheme we consider, $\sin^2 \theta_W$ is given by
$\sin^2 \theta_W\equiv 1- \frac{M_W^2}{M_Z^2}$, and this
expression is valid beyond tree--level.\\
In the on--shell case it can be shown that the interference term 
$2Re{\cal M}^{0*}{\cal M}^1$, found from squaring the one loop corrected
amplitude $|{\cal M}^0+{\cal M}^1|^2$, is equal to $\Gamma_H|{\cal M}^0|^2$.
Hence the one loop corrected width $\Gamma^1_{on}$ can be written as 
$\Gamma^1_{on}=(1+\Gamma_H)\Gamma^0_{on}$, with $\Gamma_H$ being interpreted
as the fractional contribution to the tree--level width. In the
off--shell case, and taking the  final state fermions to be massless,
 $2Re{\cal M}^{0*}{\cal M}^1$ is again equal to 
$\Gamma_H|{\cal M}^0|^2$, although $\Gamma_H$ now has a dependence on
$E_1$ and $E_2$ and thus cannot be factorized out of the phase space
integral. Therefore we define the fractional contribution to
the tree--level width as $\delta\Gamma_{off}$, with: 
$$\Gamma^1_{off}=(1+\delta\Gamma_{off})\Gamma^0_{off}$$
 Since $\Gamma_W$ does not contribute to the corrected matrix
element it is evident that the $W^+G^+$ mixing  has a 
vanishing contribution and is given in Section 3.3 for completeness.

We now briefly consider the constraints on the masses of the Higgs bosons
that can be extracted from current precision measurements of $\rho^0$, defined
by: 
\begin{equation}
\rho^0={M^2_W\over \rho M_Z^2\cos^2\theta_W}
\end{equation}
Here $\rho$ in the denominator contains all purely SM radiative
corrections, while $\rho^0 \equiv 1$ in the absence of new physics.
 In the 2HDM there are extra contributions to $\rho^0$
\cite{Rhoparam} and Ref. \cite{Langacker} shows that $-0.0017\le
 \delta\rho^0\le
0.0027$ at the $2\sigma$ level. Imposing this condition and using the formulae
in Ref. \cite{Rhoparam} we plot in Fig. 3  the allowable values of
$M_{H^{\pm}}$ and $M_A$. We vary all Higgs masses up to 500 GeV and respect
 the current experimental lower limits for 5000 randomly chosen values. 
 In Fig. 3 the triangles (points) disallow (allow) the decay 
 $H^{\pm}\to AW^{*}$. From the figure we can clearly see that
for $M_{H^{\pm}}\ge 100$ GeV the vast majority of the allowed parameter
space satisfies $M_{H^{\pm}}\ge M_A$, thus implying that
the decay $H^{\pm}\to AW^{(*)}$ will be open for $M_{H^{\pm}}$
of interest at the LHC and the Tevatron. For
$M_{H^{\pm}}\le 100$ GeV (i.e. the LEP--II range) it is easier to find
$M_{H^{\pm}}\le M_A$.

\subsection*{5.1 On--shell W gauge boson}
We now present our results for the case of the $W$ boson being on--shell.
There are three unknown parameters which determine the magnitude 
of the one loop corrected width $\Gamma_{on}^1$: $M_{H^{\pm}}$, 
$M_A$ and $\tan\beta$.
This is in contrast to the decay $H^{\pm}\to hW$ in which the
mixing angle $\alpha$ and the mass of the heavier CP--even Higgs Boson ($H$)
enter the calculation \cite{santos}. We stress that this latter analysis 
only considered the 
top--bottom  loops, while we include all the fermions corrections
and find that the light fermion loops are not entirely negligible. Moreover,
there can be significant interference among the various contributions, both
destructive and constructive. We consider both Model I and Model II, which
have effectively identical results at small $\tan\beta$, although
differ at large $\tan\beta$.

Let us discuss first the effect of a relatively light charged Higgs
 ($M_{H^{\pm}}<250$ GeV)
 and a very light CP--odd ($M_A\approx 35 $ GeV)
on $\Gamma_H$. In Fig.~4  we plot $\Gamma_H$ in Model~II as a function of 
$M_{H^{\pm}}$ for several values of $\tan\beta$. We note first that for a fixed value of
  $\tan\beta$, $\Gamma_H$ is insensitive to the variation in $M_A$
when $M_{H^{\pm}}$ is varied from 120 to 260 GeV. The peaks correspond to the
 opening of the decay
$H^+ \to \bar{t}b $.
For small  $\tan\beta$ and $M_{H^{\pm}}<170$ GeV the correction is 
rather small
($\approx 2\%$); when $M_{H^{\pm}}> 180$ GeV one can reach a correction 
of 10\%.
In the case where $\tan\beta$ is large, the effect comes exclusively from the
bottom quark mass and is around 10 \%. 

In Fig.~5 we plot $\Gamma_H$ as a function of $M_A$, 
taking $M_{H^{\pm}}=570$ GeV and 3 small values of $\tan\beta$. Since we are not
considering large $\tan\beta$ this plot is relevant for both Model~I and II.
For $M_A$ less than 300 GeV or heavier than 360 GeV the effect is about 5\%.
When  $M_A$ becomes close to $2 m_t$ a sharp peak appears and this
corresponds to the opening of the channel $A^0\to t\bar{t}$, the maximal effect
in this case being around 50\%. For $M_A$  
away from this threshold value ($M_A\approx 330 \to 345$ GeV) and for small $\tan\beta$
one can have a correction of about $-14\% \to -41\%$ . 
As $\tan\beta$ increases one quickly approaches a horizontal line at $3.3\%$. 
These effects are explained as follows: the $ttb$ loop correction is 
proportional
to $Y_{uu}$ and dominates the $bbt$ loop correction at small $\tan\beta$ 
 because
$m_t\gg m_b$ . Since $Y_{uu}$ is proportional to $1/\tan\beta$ we can explain
the $\tan\beta$ dependence in Fig. 5. As $\tan\beta$ increases the  
contribution of the $ttb$ loop weakens rapidly and the dominant 
contribution to the corrected width becomes that of the renormalized 
$e$ and $s_W$, giving a fixed value
of $\Gamma_H\approx 3.3\%$ which is very insensitive to $\tan\beta$
(note that the $bbt$ loop in Model II is proportional to $\tan\beta$ --
 see below).
We do not notice an obvious correlation between $M_{H^{\pm}}$ and $\Gamma_H$;
for the optimal case considered of $\tan\beta=0.5$ and $M_A\approx 330 $ GeV, 
varying
$M_{H^{\pm}}$ from $450$ GeV to $800$ GeV causes $\Gamma_H$ to fall
from $-18\%$ to $-27\%$.

In Fig.~6 we plot $\Gamma_H$ in Model II as a function of $\tan\beta$ for 
$\tan\beta\ge 20$. In Model I all the fermion loops decouple as $\tan\beta$
increases and one has $\Gamma_H\approx 3.3\%$ for $\tan\beta\ge 4$.
In Model II the $bbt$ loop dominates with increasing $\tan\beta$ and 
for $\tan\beta\ge 20$ the value of $\Gamma_H$ starts to differ from the 
corresponding value in Model I. Again one can find sizeable negative 
corrections, with the 
largest occurring for smaller $M_A$ i.e. the closer $M_A$ is to 
$2m_b$, the more on--shell the virtual $b$ quarks are.

In Fig.~7 we show graphically the relative magnitude of the sum of the
heavy quark loops, $ttb$ and $bbt$, compared to the sum of the
remaining fermion loops ($\Gamma_{light}$). Since we plot only
low values of $\tan\beta$ the $ttb$ contribution dominates the $bbt$
loop and so we label the sum of the $ttb$ and $bbt$ contributions as
$\Gamma_{ttb}$.
One can see that $\Gamma_{light}$ is of comparable strength
to the heavy quark loops unless $M_A$ is close to $2m_t$. In addition
there can be constructive or destructive interference, which is shown in
Fig.~7 by the sign of the ratio. 
 
\subsection*{5.2 Off--shell W gauge boson}
 We now consider the case of the W gauge boson being off--shell. Since the
decay $H^{\pm}\to AW^*$ is possible for a light $H^{\pm}$ in range at LEP--II
we shall present results for $M_{H^{\pm}}=80$ GeV, which is also in the
mass region considered problematic for detection channels which make use of
the conventional decays $H^{\pm}\to \tau\nu_{\tau}$, $cs$.
 As is mentioned in the introduction,
charged Higgs bosons of Model II are excluded from the LEP--II discovery range
by precision measurements of $b\to s\gamma$. Our discussion will therefore
be focussed on Model I.
In the massless fermion final state limit, the $WW$ self--energy is the 
only additional contribution to the one loop corrected width 
for the off--shell decay\footnote{Note that for the W being off--shell,
 there are extra contributions
coming from box diagrams which will be considered in ref. \cite{boxes}}. 
The $WW$ self--energy is the standard diagram
 and does not depend 
on $\tan\beta$. Hence all the $\tan\beta$ dependence is contained 
in the vertex contribution and in the case of Model~I is enhanced when 
$\tan\beta$ is small. 

In Fig.~8 we plot the magnitude of the one loop
corrections, $\delta\Gamma_{off}$, 
as a function of small $\tan\beta$ for two values of $M_A$. We can see
that for $\tan\beta\ge 2$ one approaches a fixed value ($\approx 2\%$)
for $\delta\Gamma_{off}$ -- this is to be interpreted (as before) as the fermion loops
decoupling, leaving a $\tan\beta$ independent value which comes from the 
$WW$ self--energy and from the renormalized $e$ and $s_W$ in the vertex 
contribution counterterms. For low $\tan\beta$ the one loop corrections
are pulled negative. Very large corrections of up to
$-90\%$ are possible for exceptionally small ($\approx 0.1$)
values of $\tan\beta$, although such values are strongly disfavoured
by measurements of $R_b$ which require $\tan\beta\ge 1.8$ ($95\%$ c.l)
for $M_{H^{\pm}}=85$ GeV \cite{Giudice}.

\section*{6.~Conclusions}
We have computed the Yukawa coupling corrections to the decay 
$H^+ \to A^0 W^+$
in the case of an on--shell and off--shell W gauge boson. We have included 
in our 
analysis both top--bottom contributions and light fermion contributions,
the latter being non--negligible and may interfere destructively or
 constructively with the former. Restrictions on the possible values
 of the Higgs boson masses from considering the $\rho$ parameter 
were also included and found to give in the majority of the cases 
$M_{H^\pm} > M_A$. In the on--shell case,
we studied the sensitivity of the Yukawa corrections to $\tan\beta$, and
 found similar effects
for small $\tan\beta$ in both Model I and Model II which can reach 50\%
for $m_A\approx 2 m_t$. 
For large $\tan\beta$, in Model I all the fermions corrections 
decouple and reach a constant value 3.3\% for $\tan\beta > 4$; 
in Model II, the top mass effect is suppressed while the bottom mass
effect is increased for $\tan\beta >20$, allowing sizeable corrections
of $10\%$ or greater. For the case of the W gauge boson being off--shell, 
the charged Higgs bosons in the LEP--II range and $\tan\beta$ not too small, 
the corrections are rather small and do not surpass 2\%.

\section*{Acknowledgements}
We thank A. Djouadi for his critical reading of the present manuscript.
A. Akeroyd was supported by DGICYT under grants PB95-1077, by the
TMR network grant ERBFMRXCT960090 of the European Union, and by a
CSIC--UK Royal Society fellowship. A.~Arhrib is very grateful to the
Laboratoire de Physique  Math\'ematique et Th\'eorique de Montpellier 
for their kind hospitality during his visit
 where part of this work has been done. 
\renewcommand{\theequation}{A.\arabic{equation}}
\setcounter{equation}{0}

\section*{Appendix A: Passarino Veltman functions }
Let us recall the definitions of scalar and tensor integrals 
\cite{passarino} we use:
\subsection*{A.1 One point function:}
The one point function is defined by:
$$ A_0(m_0^2)=
\frac{(2\pi \mu)^{4-d}}{i \pi^2}\int d^dq \frac{1}{ [q^2-m_0^2] } $$
$\mu$ is an arbitrary renormalization scale.
\begin{eqnarray}
A_0(m_0^2)= m_0^2 \left[ 1+ \Delta_0 \right] +{\cal O}(d-4)
\end{eqnarray}
The  UV divergences are contained in  $\Delta_0$  which is given by
\begin{eqnarray}
\Delta_i = \frac{2}{4-d} - \gamma_E + \log (4 \pi) +
\log \frac{\mu^2}{m_i^2}
\end{eqnarray}
note that in dimensional regularization $A_0(0)=0.$
\subsection*{A.2 Two point functions:}
The two points functions are defined by:
$$ B_{0,\mu,\mu\nu}(p_1^2,m_0^2,m_1^2)=
\frac{(2\pi \mu)^{4-d}}{i \pi^2}\int d^nq 
\frac{1, q_\mu, \mu\nu}{[q^2-m_0^2][(q+p_1)^2-m_1^2] } $$
\begin{equation}
B_{0}=\frac{1}{2}(\Delta_0 +\Delta_1) - \int_{0}^1 dx Log \frac{x^2 k^2
-x (k^2-m_0^2 + m_1^2) + m_1^2 -i \epsilon }{m_0m_1} 
\end{equation}
the derivative of $B_0$ function is defined as:
$$B_0^{'}[X, m_1^2, m_2^2]= \frac{\partial}{\partial p^2}
B_0[p^2, m_1^2, m_2^2]\vert_{p^2=X} $$
note that  $A_0$ can be expressed in term of $B_0$ 
$$ A_0(m^2) = m^2 + m^2 B_0(0,m^2,m^2) $$ using Lorentz invariance, we have:
\begin{eqnarray}
& & B_{\mu}={p_1}_{\mu} B_1 \nonumber \\
& & B_{\mu\nu}={p_1}_{\mu}{p_1}_{\mu} B_{21} + g_{\mu\nu} B_{22}\nonumber
\end{eqnarray}
\subsection*{A.3 Three point functions:}
The three point functions are defined as:
$$ C_{0,\mu,\mu \nu} (p_1^2,p_{12}^2,p_2^2,m_0^2,m_1^2,m_2^2)=
\frac{1}{i \pi^2}\int d^nq 
\frac{ 1, q_{\mu},q_{\mu}q_{\nu}}{[q^2-m_0^2][(q+p_1)^2-m_1^2]
[(q+p_2)^2-m_2^2]   } $$
where $p_{12}^2=(p_1+p_2)^2 $.
Using Lorentz invariance, $C_\mu$ and $C_{\mu\nu}$ can be written as:
\begin{eqnarray}
& & C_\mu  =  {p_1}_{\mu} C_1 +  {p_2}_{\mu} C_2 \\ [0.4 cm]
& & C_{\mu\nu}  =  g_{\mu\nu} C_{00} + {p_1}_{\mu} {p_1}_{\nu} C_{11}
+ {p_2}_{\mu} {p_2}_{\nu} C_{22} +
({p_1}_{\mu} {p_2}_{\nu} +{p_2}_{\mu} {p_1}_{\nu}) C_{12}
\end{eqnarray}

\renewcommand{\theequation}{B.\arabic{equation}}
\setcounter{equation}{0}

\section*{Appendix B: Renormalization constants}
Hereafter we give all the renormalization constants necessary to compute the
counterterms defined in eqs \ref{deltas}:
\subsection*{B.1 gauge bosons self--energies}
Let $ie \gamma^\mu (V_L \frac{1-\gamma_5}{2} + V_R \frac{1+\gamma_5}{2})$
 the general coupling of the gauge bosons  
$V_\mu$ to a pair of fermions $f$ 
and $f'$. The coefficient of $-g_{\mu\nu}$ of the self--energy of the 
gauge boson $V_\mu$ is given by:  
\begin{eqnarray}
\Sigma_T^{VV}(p^2)& = &
-\frac{N_C\alpha}{2 \pi} \{ ({V_L}^2 + {V_R}^2) [ A_0(m_{f'}^2) -
2 B_{22}(p^2, m_{f'}^2, m_f^2) + p^2 B_1(p^2, m_{f'}^2, m_f^2)
] \nonumber\\ & & +
m_f[ m_f({V_L}^2 + {V_R}^2)  
       -2 m_{f'} V_L V_R ]  B_0(p^2, m_{f'}^2, m_{f}^2)  \}
\end{eqnarray}
$\bullet$ $V=Z$,  $f=f'$, $Z_L =-\frac{1}{ s_W c_W} (T_f- e_f s_W^2)$
$Z_R =\frac{e_f s_W^2}{ s_W c_W}$, with $T_u=\frac{1}{2}$ and 
$T_d=-\frac{1}{2}$.\\
$\bullet$ $V=\gamma$, $f=f'$,  $\gamma_L=\gamma_R=-e_f$\\
$\bullet$ $V=W$, $f=u$ and $f'=d$ $W_L =-\frac{1}{\sqrt{2} s_W}$, $W_R =0$.

The renormalization constant of the electric charge is given by:
\begin{eqnarray}
 \delta Z_e &=& -\frac{1}{2} \delta Z_{\gamma\gamma}  = 
\frac{1}{2} \frac{\partial
\Sigma_T^{\gamma\gamma}(k^2)}{\partial k^2}\vert_{k^2=0}
\nonumber \\ &  =  & 
 \frac{1}{2}\frac{N_C\alpha}{3\pi} \{ e_d^2  [ -\frac{1}{3} +  
B_0(0, m_d^2, m_d^2) + 
        2 m_d^2 B_0^{'}(0, m_d^2, m_d^2)] \\ & & 
+  e_u^2 [ -\frac{1}{3} +  B_0(0, m_u^2, m_u^2) + 
        2 m_u^2 B_0^{'}(0, m_u^2, m_u^2)]\} \nonumber
\end{eqnarray}
The mass counterterms for the gauge boson $W$ and $Z$ are given by:
\begin{eqnarray}
\delta M_W^2 =\Sigma_T^{WW}(p^2=M_W^2) \qquad , \qquad 
\delta M_Z^2 =\Sigma_T^{ZZ}(p^2=M_Z^2)
\end{eqnarray}

\subsection*{B.2 Wave functions renormalization}
The wave function renormalization constants for the $W$ gauge boson can be 
obtained from the self--energy as:
\begin{eqnarray}
\delta Z_W &=& -\frac{\partial \Sigma_T^{WW}(k^2)}{\partial k^2}
\vert_{k^2=M_W^2}\nonumber \\
 &=&\frac{\alpha}{4 \pi M_W^2 s_W^2}\{
 M_W^2/3 + 
     (m_d^2 - m_u^2)^2 B_0(0,m_d^2,m_u^2)/M_W^2 \nonumber\\ & -  &
      (m_d^2 -  m_u^2)^2 + 
     M_W^4) B_0(M_W^2,m_d^2,m_u^2)/M_W^2  \nonumber\\ &  + &
 [ (m_d^2 - m_u^2)^2 + m_u^2 M_W^2 - 2 M_W^4 + 
 -2 m_d^2 M_W^2 ] B_0'(M_W^2,m_d^2,m_u^2)\}
\end{eqnarray}
The renormalization constants of the charged Higgs, 
CP--odd Higgs and the Goldstone boson wave function are given by:
\begin{eqnarray}
& & \delta Z_{H^+ H^+} = \frac{N_C \alpha}{4\pi} (
-({Y^L_{ud}}^2 + {Y^R_{ud}}^2) B_0(M_{H{\pm}}^2, m_d^2, m_u^2)  \nonumber\\ & &
 + ( [m_d^2 + m_u^2 - M_{H^{\pm}}^2 ] ( {Y^L_{ud}}^2 + {Y^R_{ud}}^2 ) + 
4 m_d m_u {Y^L_{ud}}{Y^R_{ud}} ) B'_0(M_{H^{\pm}}^2, m_d^2, m_u^2) )\nonumber \\
[0.2cm] 
& & \delta Z_{A^0} = - N_C \frac{\alpha}{2 \pi} \{ Y_{dd}^2 [ 
B_0(M_A^2,m_d^2,m_d^2) + M_A^2 B^{'}_0(M_A^2,m_d^2,m_d^2) ] +\nonumber\\ & &
Y_{uu}^2 [ 
B_0(M_A^2,m_u^2,m_u^2) + M_A^2 B^{'}_0(M_A^2,m_u^2,m_u^2) ] \}  \\
[0.2cm]
& & \delta Z_{G^+ G^+} = \frac{\alpha N_C}{8 \pi M_W^2 s_W^2}\{
 -(m_d^2+m_u^2) B_0(0,m_d^2,m_u^2)+ 
         (m_d^2 - m_u^2)^2 B^{'}_0(0,m_d^2,m_u^2)\} \nonumber
\end{eqnarray}

\newpage

\subsection*{Figure Captions}

\vspace*{0.5cm}

\renewcommand{\labelenumi}{Fig. \arabic{enumi}}
\begin{enumerate}

\item  
Feynman diagrams for the Born approximation to the decay 
$H^+ \to A^0 W^{+*}$, W on--shell (1.a), W off--shell (1.b).

\item  
Feynman diagrams for the one loop corrections to the decay 
$H^+ \to A^0 W^{+*}$:  vertex (2.$a_1$ and 2.$a_1$), WW self--energy (2.$a_3$),
W--G mixing (2.$a_4$). Charged Higgs boson, Goldstone boson and CP--odd
 self--energies (2.$b_{1,2,3}$). (2.$b_{4,5,6}$) WW, ZZ and $\gamma\gamma$
self--energies.  (2.$c_{1}$) is the vertex counterterm, (2.$c_{2}$)
W self--energy counterterm and (2.$c_{5}$) is the $W$--$G$ mixing  counterterm.

\vspace{5mm}
\item  
Scatter plot of values of $M_{H^\pm}$ and $M_A$ 
consistent with measurements of $\rho^0$. 
Triangles disallow the decay $H^{\pm}\to AW^{(*)}$.

\vspace{5mm}
\item  
$\Gamma_H$ as a function of $M_{H^{\pm}}$ (Model II)
for $M_A=35$ GeV, $\tan\beta=0.5$, $1.0$, $4$ and 70.

\vspace{5mm}
\item  
$\Gamma_H$ as a function of $M_A$ (Models I and II)
for $\tan\beta=0.6$, $1.0$ and 2.5.

\vspace{5mm}
\item  
$\Gamma_H$ as a function of $\tan\beta$ (Model  II)
for $\tan\beta\ge 20$.

\vspace{5mm}
\item  
$\Gamma_{ttb}/\Gamma_{light}$ 
as a function of small $\tan\beta$ for several values of $M_A$
(Models I and II). 

\vspace{5mm}
\item  
$\delta\Gamma_{off}$  
as a function of small $\tan\beta$ and for $M_A=15$, $40$ GeV (Model I).

\end{enumerate}

\newpage

\vspace{-0.8cm}
\begin{picture}(100,100)(0,0)
\put(51,0) {\hpe{$H^\pm$}{10}}
\put(88,0){\line(1,1){28}}
\put(88,0){\line(1,-1){28}}
\put(117,-27){\line(0,1){51}}
\put(116,25){\circle*{7}}
\put(116,-25){\circle*{7}}
\put(88,0){\circle*{5}}
\put(92,16){\makebox{$u$}}
\put(89,-19){\makebox{$d$}}
\put(119,0){\makebox{$u$}}
\put(116,25) {\hpe{$A^0$}{10}}
\multiput(120,-25)(16,0){2}{\oval(8,8)[t]}
\multiput(128,-25)(16,0){2}{\oval(8,8)[b]}
\put(99,11){\vector(-1,-1){1}}
\put(117,0){\vector(0,1){1}}
\put(101,-13){\vector(1,-1){1}}
\put(127,-17){\makebox{$W^\pm$}}
\put(101,-45){\makebox{Fig.2.$a_1$}}
\put(258,0) {\hpe{$H^\pm$}{10}}
\put(295,0){\line(1,1){28}}
\put(295,0){\line(1,-1){28}}
\put(324,-27){\line(0,1){51}}
\put(323,25){\circle*{7}}
\put(323,-25){\circle*{7}}
\put(295,0){\circle*{5}}
\put(299,16){\makebox{$d$}}
\put(296,-19){\makebox{$u$}}
\put(326,0){\makebox{$d$}}
\put(323,25) {\hpe{$A^0$}{10}}
\multiput(327,-25)(16,0){2}{\oval(8,8)[t]}
\multiput(335,-25)(16,0){2}{\oval(8,8)[b]}
\put(306,11){\vector(-1,-1){1}}
\put(324,0){\vector(0,1){1}}
\put(308,-13){\vector(1,-1){1}}
\put(335,-17){\makebox{$W^\pm$}}
\put(301,-45){\makebox{Fig.2.$a_2$}}
\end{picture}

\vspace{1.9cm}

\begin{picture}(100,100)(0,0)
\put(57,50) {\unfsp{$A^0$}}
\multiput(12,50)(8,0){6}{\line(1,0){6}}
\put(12,54){\mbox{$H^+$}}
\put(61,50){\circle*{5}}
\multiput(65.8,50)(16,0){2}{\oval(8,8)[t]}
\multiput(74,50)(16,0){2}{\oval(8,8)[b]}
\put(105,50){\circle{21}}
\multiput(121,50)(16,0){2}{\oval(8,8)[t]}
\multiput(129,50)(16,0){2}{\oval(8,8)[b]}
\put(101,62){\makebox{$u$}}
\put(102,29){\makebox{$d$}}
\put(78,54){\makebox{$W^\pm$}}
\put(123,54){\makebox{$W^\pm$}}
\put(95,48){\circle*{5}}
\put(115,51){\circle*{5}}
\put(100,8){\makebox{Fig.2.$a_3$}}
\put(287,50) {\unfsp{$A^0$}}
\multiput(242,50)(8,0){7}{\line(1,0){6}}
\put(242,54){\mbox{$H^+$}}
\put(291,50){\circle*{5}}
\multiput(295,50)(10.8,0){3}{\line(1,0){5.2}}
\put(335,50){\circle{21}}
\multiput(351,50)(16,0){2}{\oval(8,8)[t]}
\multiput(359,50)(16,0){2}{\oval(8,8)[b]}
\put(331,62){\makebox{$u$}}
\put(332,29){\makebox{$d$}}
\put(308,54){\makebox{$G^\pm$}}
\put(353,54){\makebox{$W^\pm$}}
\put(325,50){\circle*{5}}
\put(345,51){\circle*{5}}
\put(305,8){\makebox{Fig.2.$a_4$}}
\end{picture}

\vspace{.2cm}

\begin{picture}(100,100)(0,0)
\put(19,50) {\hpe{$H^\pm$}{10}}
\put(50.3,50){\circle*{5}}
\put(65,74){\makebox{$u$}}
\put(65,21){\makebox{$d$}}
\put(70,50){\circle{35}}
\put(87,50) {\hpe{$H^\pm$}{20}}
\put(88,50){\circle*{5}}
\put(68,68){\vector(-1,0){1}}
\put(71,32){\vector(1,0){1}}
\put(60,3){\makebox{Fig.2.$b_1$}}
\put(149,50) {\hpe{$G^\pm$}{10}}
\put(180.3,50){\circle*{5}}
\put(195,74){\makebox{$u$}}
\put(195,21){\makebox{$d$}}
\put(200,50){\circle{35}}
\put(217,50) {\hpe{$G^\pm$}{20}}
\put(218,50){\circle*{5}}
\put(198,68){\vector(-1,0){1}}
\put(201,32){\vector(1,0){1}}
\put(189,3){\makebox{Fig.2.$b_2$}}
\put(269,50) {\hpe{$A^0$}{10}}
\put(300.3,50){\circle*{5}}
\put(315,74){\makebox{$u,d$}}
\put(315,21){\makebox{$u,d$}}
\put(320,50){\circle{35}}
\put(337,50) {\hpe{$A^0$}{20}}
\put(338,50){\circle*{5}}
\put(318,68){\vector(-1,0){1}}
\put(321,32){\vector(1,0){1}}
\put(310,3){\makebox{Fig.2.$b_3$}}
\end{picture}

\vspace{.2cm}

\begin{picture}(100,100)(0,0)
\put(62,66){\vector(-1,0){1}}
\put(66,34){\vector(1,0){1}}
\put(49.6,50){\circle*{4}}
\put(80.3,50){\circle*{4}}
\multiput(21,50)(16,0){2}{\oval(8,8)[t]}
\multiput(29,50)(16,0){2}{\oval(8,8)[b]}
\put(65,50){\circle{30}}
\multiput(85,50)(16,0){2}{\oval(8,8)[t]}
\multiput(93,50)(16,0){2}{\oval(8,8)[b]}
\put(56,72){\makebox{$u$}}
\put(60,23){\makebox{$d$}}
\put(20,60){\makebox{$W^\pm$}}
\put(92,60){\makebox{$W^\pm$}}
\put(49,50){\circle*{5}}
\put(81,51){\circle*{5}}
\put(60,3){\makebox{Fig.2.$b_4$}}
\put(207.8,65.8){\vector(-1,0){1}}
\put(211,34){\vector(1,0){1}}
\put(194.6,50){\circle*{4}}
\put(225.3,50){\circle*{4}}
\multiput(166,50)(16,0){2}{\oval(8,8)[t]}
\multiput(174,50)(16,0){2}{\oval(8,8)[b]}
\put(211,50){\circle{30}}
\multiput(230,50)(16,0){2}{\oval(8,8)[t]}
\multiput(238,50)(16,0){2}{\oval(8,8)[b]}
\put(201,72){\makebox{$u,d$}}
\put(205,23){\makebox{$u,d$}}
\put(165,60){\makebox{$ Z$}}
\put(237,60){\makebox{$ Z$}}
\put(194,50){\circle*{5}}
\put(226,51){\circle*{5}}
\put(194,3){\makebox{Fig.2.$b_5$}}
\put(327,66){\vector(-1,0){1}}
\put(331,34){\vector(1,0){1}}
\put(314.6,50){\circle*{4}}
\put(345.3,50){\circle*{4}}
\multiput(286,50)(16,0){2}{\oval(8,8)[t]}
\multiput(294,50)(16,0){2}{\oval(8,8)[b]}
\put(330,50){\circle{30}}
\multiput(350,50)(16,0){2}{\oval(8,8)[t]}
\multiput(358,50)(16,0){2}{\oval(8,8)[b]}
\put(321,72){\makebox{$u,d$}}
\put(325,23){\makebox{$u,d$}}
\put(300,60){\makebox{$\gamma$}}
\put(357,60){\makebox{$\gamma$}}
\put(314,50){\circle*{5}}
\put(346,51){\circle*{5}}
\put(314,3){\makebox{Fig.2.$b_6$}}
\end{picture}

\vspace{.2cm}

\begin{center}
\begin{picture}(100,100)(0,0)
\put(-90,52) {\unbs{$W^+$}}
\put(-90,51) {\unfsp{$A^0$}}
\multiput(-140,50)(8,0){7}{\line(1,0){6}}
\put(-131,54){\mbox{$H^+$}}
\put(-94,44){\mbox{{\Huge +}}}
\put(-99,10){\mbox{{Fig.2.$c_1$}}}
\multiput(30,50)(16,0){2}{\oval(8,8)[t]}
\multiput(38,50)(16,0){2}{\oval(8,8)[b]}
\multiput(65,50)(16,0){2}{\oval(8,8)[t]}
\multiput(73,50)(16,0){2}{\oval(8,8)[b]}
\put(51,44){\mbox{{\Huge +}}}
\put(25,60){\makebox{$W^\pm$}}
\put(77,60){\makebox{$W^\pm$}}
\put(38,10){\mbox{{Fig.2.$c_2$}}}
\multiput(150,50)(16,0){2}{\oval(8,8)[t]}
\multiput(158,50)(16,0){2}{\oval(8,8)[b]}
\multiput(185,50)(8,0){4}{\line(1,0){6}}
\put(177,44){\mbox{{\Huge +}}}
\put(145,60){\makebox{$W^\pm$}}
\put(197,60){\makebox{$G^\pm$}}
\put(157,10){\mbox{{Fig.2.$c_3$}}}
\end{picture}
\end{center}

\vspace{.9cm}

\centerline{\bf{Figure 2} }

\begin{minipage}[t]{19.cm}
\setlength{\unitlength}{1.in}
\begin{picture}(1,1)(.7,7.5)
\centerline{\epsffile{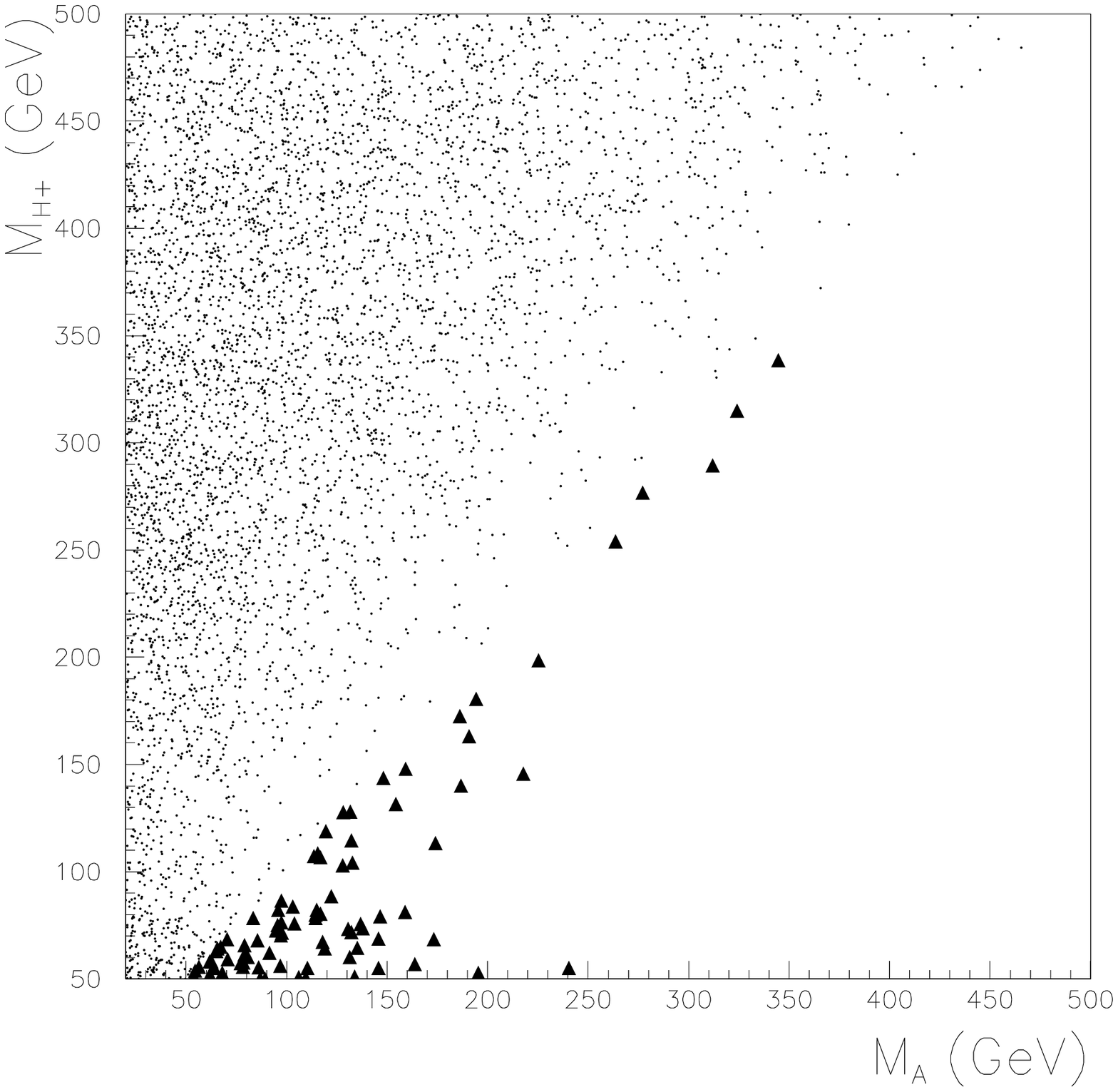}}
\end{picture}
\vspace{16.5cm}

\hspace{6.cm}\bf{Figure. 3} 
\end{minipage}

\newpage
\begin{minipage}[t]{19.cm}
\setlength{\unitlength}{1.in}
\begin{picture}(1,1)(.7,7.5)
\centerline{\epsffile{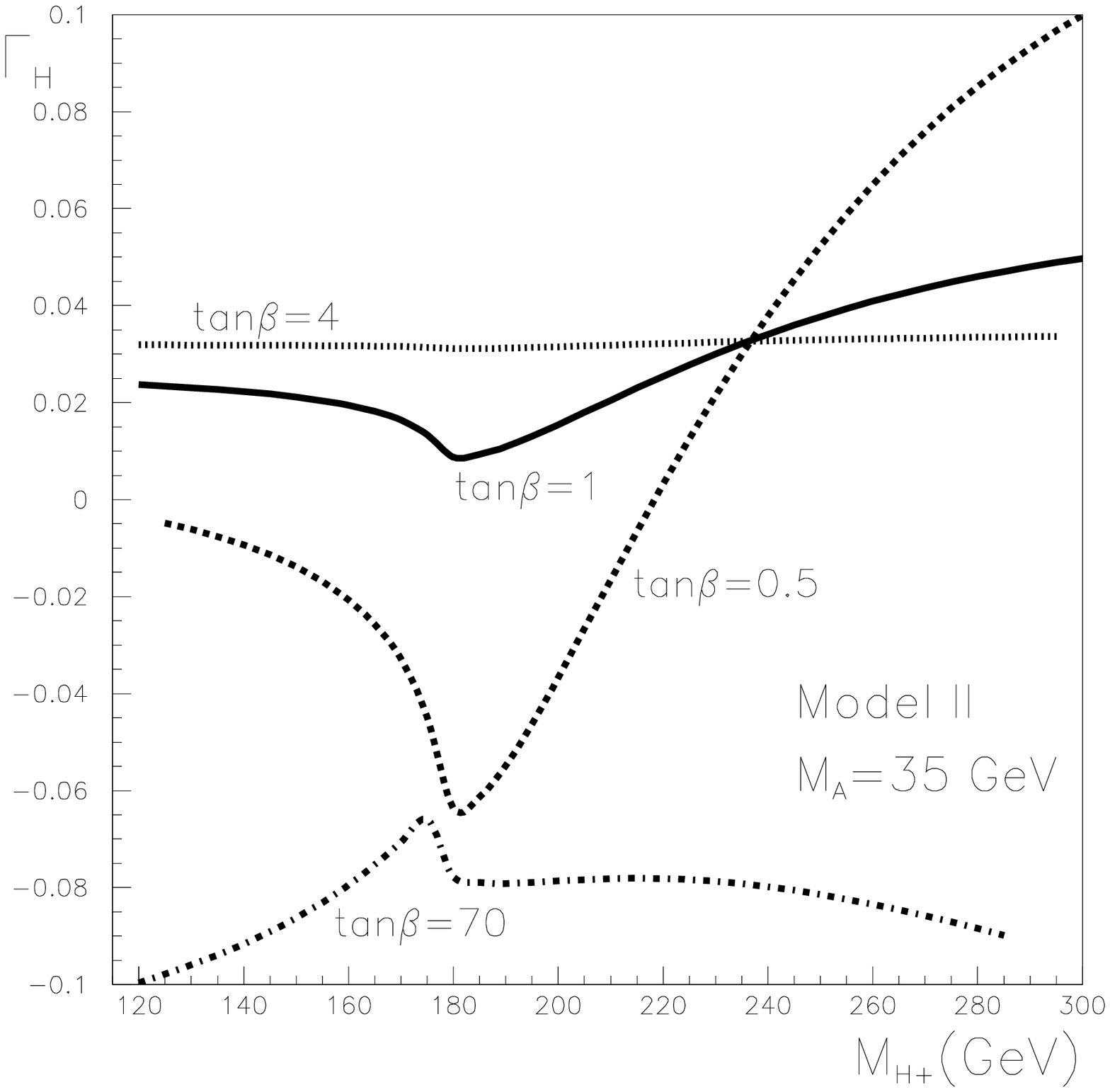}}
\end{picture}
\vspace{16.5cm}

\hspace{6.cm}\bf{Figure. 4}  
\end{minipage}

\newpage
\begin{minipage}[t]{19.cm}
\setlength{\unitlength}{1.in}
\begin{picture}(1,1)(.7,7.5)
\centerline{\epsffile{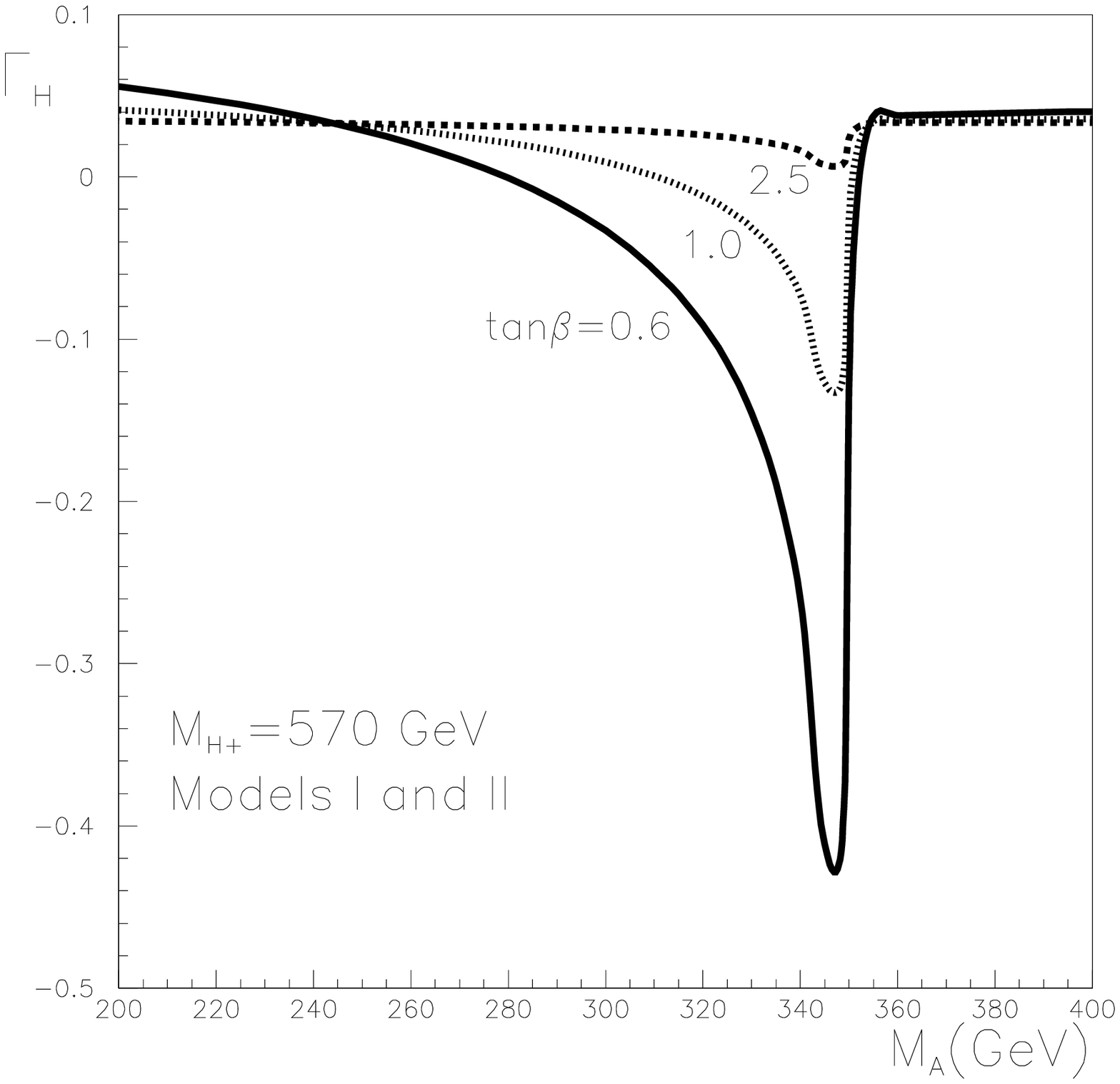}}
\end{picture}
\vspace{16.5cm}

\hspace{6.cm}\bf{Figure. 5}  
\end{minipage}

\newpage
\begin{minipage}[t]{19.cm}
\setlength{\unitlength}{1.in}
\begin{picture}(1,1)(.7,7.5)
\centerline{\epsffile{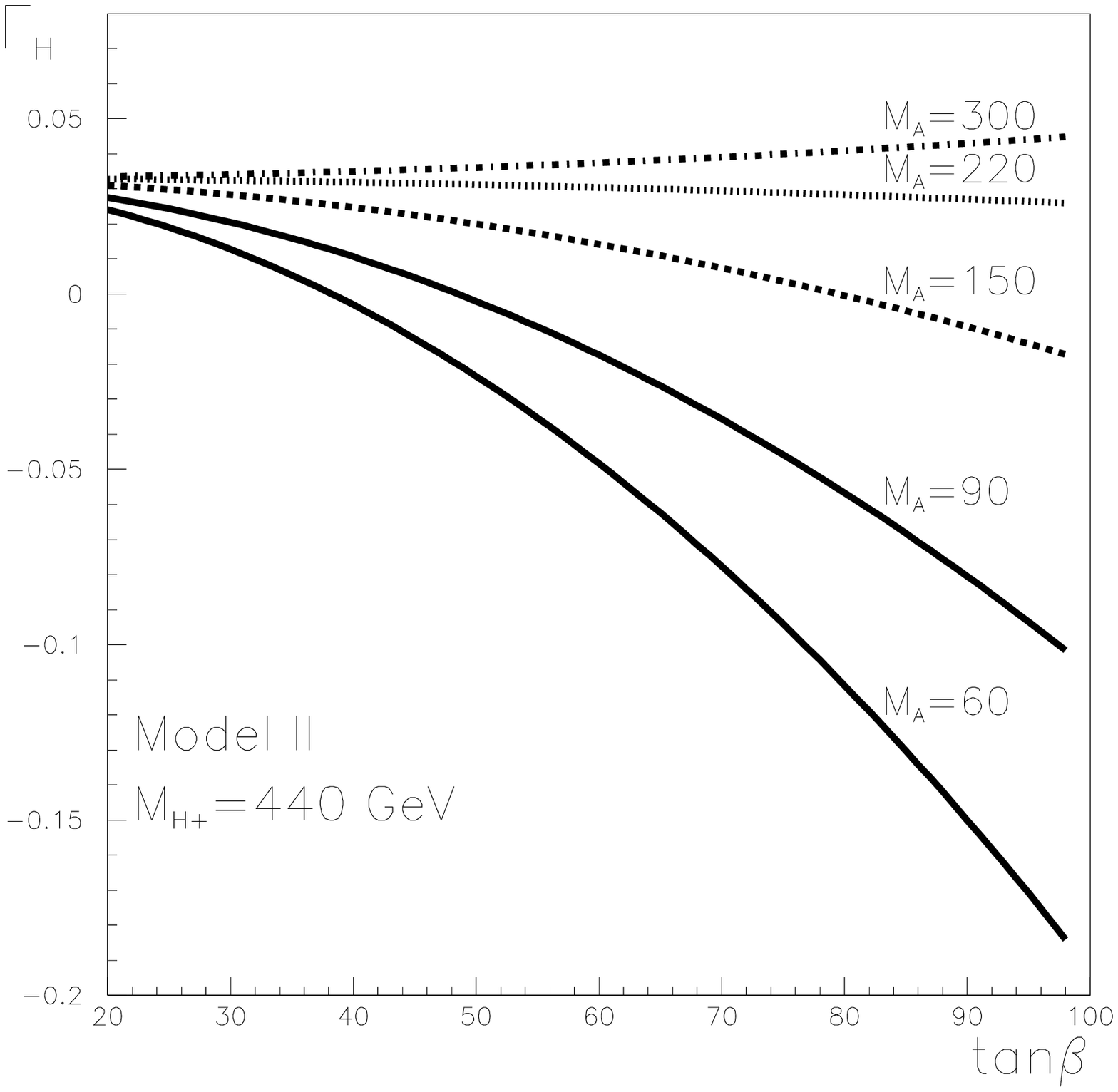}}
\end{picture}
\vspace{16.5cm}

\hspace{6.cm}\bf{Figure. 6} 

\end{minipage}

\newpage
\begin{minipage}[t]{19.cm}
\setlength{\unitlength}{1.in}
\begin{picture}(1,1)(.7,7.5)
\centerline{\epsffile{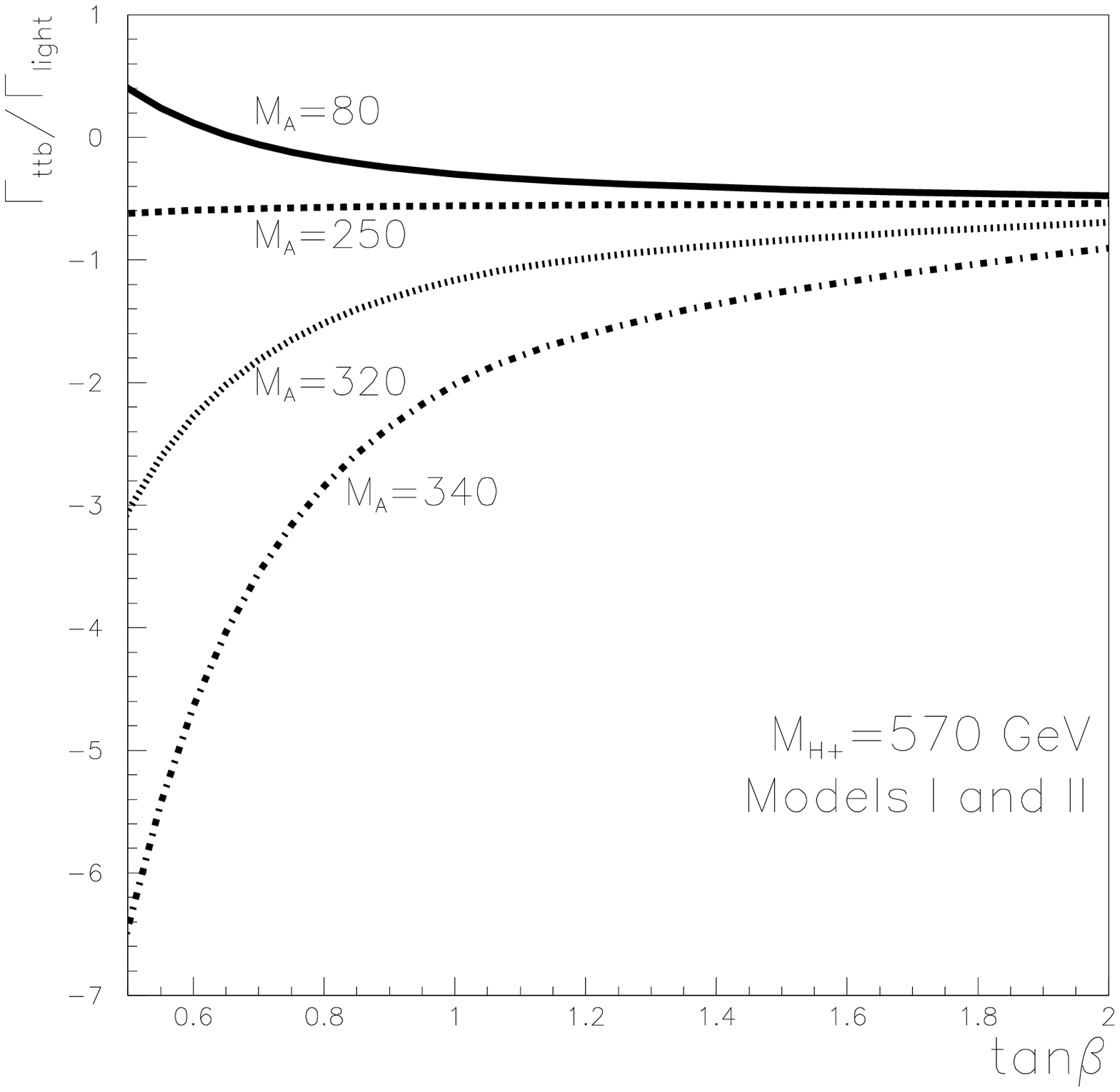}}
\end{picture}
\vspace{16.5cm}

\hspace{6.cm}\bf{Figure. 7} 
\end{minipage}

\newpage
\begin{minipage}[t]{19.cm}
\setlength{\unitlength}{1.in}
\begin{picture}(1,1)(.7,7.5)
\centerline{\epsffile{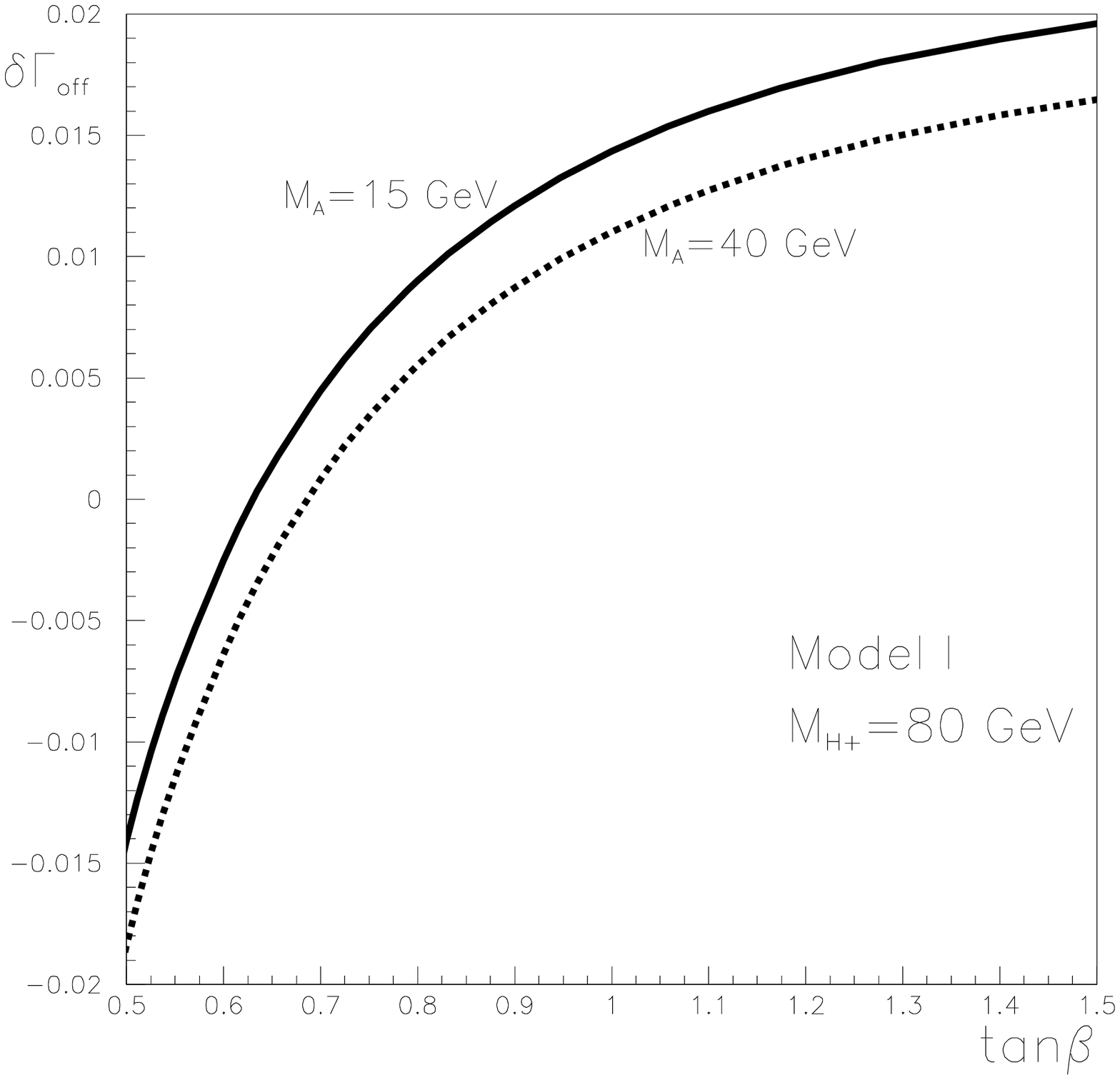}}
\end{picture}
\vspace{16.5cm}

\hspace{6.cm}\bf{Figure. 8} 
\end{minipage}

\end{document}

%% file: fey1.tex

\def\unbs#1{
\begin{picture}(100,200)(0,0)
\put(0,0){\circle*{5}}
\multiput(5,0)(8,-8){6}{\oval(8,8)[bl]}
\multiput(5,-8)(8,-8){6}{\oval(8,8)[tr]}
\put(37,-17){\makebox(0,0){#1}}
\end{picture}   }

\def\be#1#2{
\begin{picture}(100,200)(0,0)
\put(25,0){\circle*{5}}
\multiput(0,23)(8,-8){3}{\oval(8,8)[tr]}
\multiput(8,23)(8,-8){3}{\oval(8,8)[bl]}
\multiput(0,-23)(8,8){3}{\oval(8,8)[br]}
\multiput(8,-23)(8,8){3}{\oval(8,8)[tl]}
 
\put(-10,26){\makebox(0,0){#1}}
\put(-10,-26){\makebox(0,0){#2}}
\end{picture}   }
 
\def\bs#1#2{
\begin{picture}(100,200)(0,0)
\put(0,0){\circle*{5}}
\multiput(5,0)(8,8){3}{\oval(8,8)[tl]}
\multiput(5,8)(8,8){3}{\oval(8,8)[br]}
\multiput(5,0)(8,-8){3}{\oval(8,8)[bl]}
\multiput(5,-8)(8,-8){3}{\oval(8,8)[tr]}
 
\put(34,30){\makebox(0,0){#1}}
\put(34,-30){\makebox(0,0){#2}}
\end{picture}   }
\def\fe#1#2{
\begin{picture}(100,200)(0,0)
\put(34,0){\circle*{5}}
\put(0,35){\vector(1,-1){35}}
\put(-10,26){\makebox(0,0){#1}}
\put(-10,-26){\makebox(0,0){#2}}
\put(0,-35){\vector(1,1){35}}
\end{picture} }
%
\def\fs#1#2{
\begin{picture}(100,200)(0,0)
\put(0,0){\circle*{5}}
\put(0,0){\vector(1,1){25}}
\put(34,26){\makebox(0,0){#1}}
\put(34,-26){\makebox(0,0){#2}}
\put(0,0){\vector(1,-1){25}}
\end{picture} }
\def\bp#1#2{
\begin{picture}(100,200)(0,0)
\multiput(0,0)(16,0){3}{\oval(8,8)[t]}
\multiput(8,0)(16,0){3}{\oval(8,8)[b]}
 
\put(#2,15){\makebox(0,0){#1}}
\end{picture}   }
\def\fp#1#2{
\begin{picture}(100,200)(0,0)
\put(0,0){\line(1,0){48}}
\put(#2,15){\makebox(0,0){#1}}
\end{picture}   }
%
\def\fpe#1#2{
\begin{picture}(100,200)(0,0)
\put(0,0){\line(1,0){35}}
\put(#2,15){\makebox(0,0){#1}}
\end{picture}   }

\def\fpp#1#2{
\begin{picture}(100,200)(0,0)
\put(0,0){\line(1,0){48}}
\put(#2,15){\makebox(0,0){#1}}
\end{picture}   }
\def\ft#1{
\begin{picture}(100,200)(0,0)
\put(0,0){\line(0,1){48}}
\put(10,24){\makebox(0,0){#1}}
\end{picture}   }
\def\fth#1{
\begin{picture}(100,200)(0,0)
\put(0,0){\line(0,1){48}}
\put(10,24){\makebox(0,0){#1}}
\end{picture}   }
 
\def\bt#1{
\begin{picture}(100,200)(0,0)
\multiput(0,0)(0,16){3}{\oval(8,8)[tl]}
\multiput(0,8)(0,16){3}{\oval(8,8)[br]}
\multiput(0,8)(0,16){3}{\oval(8,8)[tr]}
\multiput(0,0)(0,16){3}{\oval(8,8)[bl]}
 
\put(15,24){\makebox(0,0){#1}}
\end{picture}   }
\def\fpp#1#2{
\begin{picture}(100,200)(0,0)
\multiput(0,0)(10.8,0){6}{\line(1,0){5.2}}
\put(#2,15){\makebox(0,0){#1}}
\end{picture}   }
\def\hpe#1#2{
\begin{picture}(100,200)(0,0)
\multiput(0,0)(10.8,0){3}{\line(1,0){5.2}}
\put(#2,8){\makebox(0,0){#1}}
\end{picture}   }

\def\ftp#1{
\begin{picture}(100,200)(0,0)
\multiput(0,0)(0,10.8){5}{\line(0,1){5.2}}
\put(10,24){\makebox(0,0){#1}}
\end{picture}   }

\def\fsp#1#2{
\begin{picture}(100,200)(0,0)
\put(0,0){\circle*{5}}
\multiput(0,0)(16,16){3}{\line(1,1){10}}
\multiput(0,0)(16,-16){3}{\line(1,-1){10}}
 
\put(27,36){\makebox(0,0){#1}}
\put(52,-36){\makebox(0,0){#2}}
\end{picture}   }

\def\fep#1#2{
\begin{picture}(100,200)(0,0)
\put(40,0){\circle*{5}}
\multiput(0,-40)(16,16){3}{\line(1,1){5}}
\multiput(0,40)(16,-16){3}{\line(1,-1){5}}
 
\put(-7,36){\makebox(0,0){#1}}
\put(-7,-36){\makebox(0,0){#2}}
\end{picture}   }
\def\unfsp#1{
\begin{picture}(100,200)(0,0)
\put(0,0){\circle*{5}}
\multiput(0,0)(16,16){3}{\line(1,1){10}}
\put(45,36){\makebox(0,0){#1}}
\end{picture}   }